# About the Heisenberg's uncertainty principle and the determination of effective optical indices in integrated photonics at high sub-wavelength regime

B. Bêche[a,b], E. Gaviot[c]

[a] Institut de Physique de Rennes - IPR UMR CNRS 6251, Université de Rennes 1, 35042 Rennes, France.
[b] Institut Universitaire de France - IUF, 75005 Paris, France.
[c] Laboratoire d'Acoustique de l'Université du Maine - LAUM UMR CNRS 6613, Université du Maine, 72000 Le Mans, France.

Email: bruno.beche@univ-rennes1.fr

**Abstract**

Within the Heisenberg's uncertainty principle it is explicitly discussed the impact of these inequalities on the theory of integrated photonics at sub-wavelength regime. More especially, the uncertainty of the effective index values in nanophotonics at sub-wavelength regime, which is defined as the eigenvalue of the overall opto-geometric problems in integrated photonics, appears directly stemming from Heisenberg's uncertainty. An apt formula is obtained allowing us to assume that the incertitude and the notion of eigenvalue called effective optical index or propagation constant is inversely proportional to the spatial dimensions of a given nanostructure yielding a transfer of the fuzziness on relevant senses of eigenvalues below a specific limit's volume.

The purpose of this letter is to discuss and study how the Heisenberg's principle acts on the notion of optical effective indices in integrated photonics at sub-wavelength regime or nanophotonics. Considering physics of guided waves [1, 2] and their applications [3] in telecommunications or sensors it is usual to resort to the theory of the electromagnetic fields in waveguides structures so as to obtain the eigenvalues equations of the whole system for hyper-frequency and photonic wavelengths. The methodology consists in solving the equations of J. C. Maxwell in every under-parts of the global system while taking into account the continuity properties of the electromagnetic fields, then eliminating the constant of integration with the limit conditions so as to obtain the so-called eigenvalues equations. In fact, the latter stand for the electromagnetic-geometric or opto-geometric equations with known-data of the system such as wavelength, permittivity or indices in optics with various sizes, is also considered a function of an unknown data in terms of eigenvalue specified by a number of integers or quantification numbers linked to the 1, 2, or 3 dimensions of the system [4, 5]. Resolution of such equations with analytical or numerical methods [6, 7] yields a series of eigenvalues called either effective propagation constants $\beta$, or effective permittivities $\varepsilon_{eff}$, or effective indices $n_{eff}$, or effective wavelength $\lambda_{eff}$, that highlights directly the overall quantifications of the fields called eigenvectors in electromagnetic modes.

$$\beta = k_0 n_{eff} = \frac{2\pi}{\lambda_0} n_{eff} = \frac{2\pi}{\lambda_{eff}} \qquad \text{with} \quad n_{eff} \equiv \sqrt{\varepsilon_{eff}} \qquad (1)$$

Typically, as the opto-geometrical system in integrated electromagnetics or nanophotonics shows off an asymmetry of pure form in either permittivities or optical indices, a cut appears in the dispersion-curves of the modes. In such cases, the electromagnetic or optical modes associated with the quantified light will not occur within the global structures, forbidding then the assessment of the family of eigen-values and -vectors. As a prominent property, in some particular symmetric cases of pure geometry in shapes and optical indices (for example: pure symmetric planar waveguides with identic lower and upper-optical-cladding, pure cylindrical or tubular shapes [8]), there no cut on the dispersion curves for specific modes whatever the scale of the system (for example the $TE_{00}$ and $TM_{00}$ single-modes into symmetric planar waveguides, the $HE_{11}$ optical modes in cylinders or tubular structures). In such cases the quantified light defined by the aforementioned modes can take place and propagate into the specific structures whatever the sub-wavelength dimension of the core waveguide (high sub-$\lambda$ regime, nanophotonics), then, no limitation occurs about the pure mathematical existence of the modes defined with (1), considering for example the determination of effective index $n_{eff}$ values at such a scale. The aim of the present note is to discuss the impact of W. Heisenberg's uncertainty principle [9, 10] on the determination of effective optical indices within symmetrical integrated photonic structures at high sub-wavelength regime.

Multiplying previous eigenvalues (1) with $\hbar = h/2\pi$, the Planck's constant allows us to define directly the well-known quantified energy and momentum properties related to the photon [11]:

$$E = \hbar\omega_{eff} = \frac{hc}{\lambda_{eff}}, \quad p = \hbar\beta = \frac{h\nu_{eff}}{c} = \frac{h}{\lambda_{eff}} = \frac{h}{\lambda_0}n_{eff} \qquad (2)$$

where c is the celerity of light in vacuum in physics, $\omega_{eff}=2\pi\nu_{eff}$ the effective pulsation of light, E the energy and p the momentum of the photon.

In both theoretical and applied physics, the concept of reciprocal spaces is defined with mathematical links of transforms called Fourier, Laplace, or z. Considering the global theory of signals, the Telegraphers equation or the linewidth equivalence, if two signals are 'trans-Fourier' from each other, then the product of their respective width exceeds a fixed quantity:

$$\Delta x . \Delta k_0 \geq \frac{1}{2}, \quad \Delta\omega . \Delta t \geq \frac{1}{2} \qquad (3)$$

Starting from such classical but crucial signal and electromagnetic field inequalities subjecting the product to a quantity standing in the second member, then multiplying per $\hbar$, the elementary action on quantum physics own account, thus yield the straight definition of the Heisenberg's uncertainty relations and Mandelstam-Tamm relation [12]:

$$\Delta x . \Delta p \geq \frac{\hbar}{2}, \text{ and } \Delta E . \Delta t \geq \frac{\hbar}{2} \qquad (4)$$

Both relations address spatial and temporal aspects which are totally equivalent since p=E/c. Then, it is clear that the quantum characteristics at the origin of the Heisenberg's relations are only due to both relations (2), aside (3). The energy of a given state will better be known as the time $\Delta t$ is increased, making then $\Delta t \leq \frac{\hbar}{2.\Delta E}$ to appear as the necessary suitable duration to measure such a state. This can also be seen as flexibility on the energy conservation in physics which is quite less strict in quantum than in classic physics.

Considering (1) prior to differentiating the second part of (2) leads to:

$$\Delta p = \hbar \Delta \beta = h \cdot \left[\frac{\Delta n_{eff}}{\lambda_0} + \frac{n_{eff}}{\lambda_0^2}\Delta\lambda_0\right] \quad \text{or} \quad \Delta\beta = 2\pi \cdot \left[\frac{\Delta n_{eff}}{\lambda_0} + \frac{n_{eff}}{\lambda_0^2}\Delta\lambda_0\right] \quad (5)$$

Let's consider now, together with Eq. (5), the textbook case of a Dirac spectral light with $\Delta\lambda_0 \to 0$, then the uncertainty relation (4) related to $\Delta x \cdot \Delta\beta \geq \frac{1}{2}$ yields relevant inequalities on the eigenvalues regarding integrated photonics with:

$$\Delta x \cdot \Delta n_{eff} \geq \frac{\lambda_0}{4\pi}, \quad \text{or} \quad \Delta x \cdot \Delta\lambda_{eff} \geq \frac{(\lambda_{eff})^2}{4\pi} \quad (6)$$

We can notice that for both these uncertainty relations on eigenvalues of integrated photonics, each member addresses a same dimension. Then, according to values of $\Delta x$ and $\Delta\lambda$ close to µm or sub-µm in nanophotonics, the dimensions of the previous inequalities range around respectively nm and nm$^2$.

Let's consider now the existence of a Full Width Half Maximum (FWHM $\equiv \Delta\lambda$) with the spectral aspect of light into relation (5). Then, this light should exhibit a coherence time $T_c=1/\Delta\nu$ during which the amplitude and the phase could be considered constant, leading to a coherence length of the light-train described as: $L_c = c \cdot T_c$, with:

$$L_c = \frac{\lambda_0^2}{\Delta\lambda} \quad (7)$$

Substituting (7) into (5) yields then the following relations:

$$\Delta p = h\left[\frac{n_{eff}}{L_c} + \frac{\Delta n_{eff}}{\lambda_0}\right] \quad \text{or} \quad \Delta\beta = 2\pi \cdot \left[\frac{n_{eff}}{L_c} + \frac{\Delta n_{eff}}{\lambda_0}\right] \quad (8)$$

Then, according to (8), the Heisenberg's uncertainty relation (4) related to the $n_{eff}$-eigenvalues on integrated photonics can be written as:

$$\Delta n_{eff} \geq \lambda_0 \cdot \left[\frac{1}{4\pi \cdot \Delta x} - \frac{n_{eff}}{L_c}\right] \quad (9)$$

One can notice that for all known materials relevant effective index values are globally ranged between 1 and 4: This can be considered either for the family of bound modes with dispersion curves located within the cone of light, or for the family of leaky modes that present a real part of the eigenvalue value close and below the optical index of claddings. The optical wavelengths dealt with nanophotonics for numerous applications are typically ranging from the visible to the infrared with 0.4µm <$\lambda_0$<10 µm. Then, with regard to relation (9) that describes the uncertainty on the eigenvalue called effective index of the modes in nanophotonics and sub-wavelength regime, the second member consists of two competing terms respectively close to ($\lambda_0/\Delta x$) and ($\lambda_0/L_c$). The coherence length $L_c$ of light sources typically used in integrated photonics (such as superluminescente diodes with large FWHM or monochromatic lasers), can be respectively assessed from ten microns up to the meter. Then, the second ($\lambda_0/L_c$)-term ranging typically between [10$^{-6}$ - 0.1] should generally not prove as much as an impact compared to the first one ($\lambda_0/\Delta x$) whose value can achieve a few hundreds up to the thousand in relation (9) with nanophotonics. Indeed, considering first a sub-wavelength regime working at a $\lambda$-µm wavelength with nanophotonic devices, then the fact that all the actual thin-layer processes developed in any clean room with e-beam lithography allow to shape optical nanometer circuits: then, such devices are clearly able to localize the light and its associated modes within a Δx close to a few nanometers scale, and soon below such a scale with symmetric structures devoid of cut-off. In such cases, the Heisenberg's uncertainty acts *via* the relation (9) showing a significant uncertainty on the value of the eigenvalues $n_{eff}$, even if one carries on with the procedures relevant to the electromagnetism wave theory which seem to work finely in such a sub-lambda regime and ultra-short dimension structures. In other words, considering a strict theoretical aspect, relation (9) highlights by itself a source of variability or fuzziness directly emerging from the Heisenberg's principle. Moreover, it confirms the optical index notion to be quite a macroscopic feature submitted to the size of a minimum volume so as to be properly defined. Such a volume corresponds to the dimension of the cell used in optics and electromagnetics for example in computing with various numerical methods such as the Beam Propagation Method (BPM) or the Finite Difference Time Domain (FDTD method)… According to (9), below such a volume limit the definition of the effective index becomes unattainable due to the increased eigenvalue uncertainty. Then, with regard to the wave nature of the light interpretation which is based on its spatial extension, we can eventually posit that the Heisenberg's principle allows for an apt limitation on the optical effective index definition below a specific volume that not allowing the definition and notion of macroscopic effective indices.